\documentclass[prl, nofootinbib, aps, twocolumn, superscriptaddress,
 preprintnumbers, amsmath, amssymb, floatfix]{revtex4}
\usepackage{amsmath}
\usepackage{amssymb}
\usepackage{graphicx}
\newcommand{\keV}{\mathrm{keV}}

\newcommand{\GeV}{\mathrm{GeV}}
\newcommand{\TeV}{\mathrm{TeV}}
\newcommand{\seconds}{\mathrm{s}}
\newcommand{\MP}{M_{\mathrm{P}}}
\newcommand{\stau}{{\widetilde{\tau}_1}}
\newcommand{\neutralino}{{\widetilde \chi}^{0}_{1}}

\newcommand{\mgr}{m_{\widetilde{G}}}
\newcommand{\mstau}{m_{\widetilde{\tau}_1}}
\newcommand{\Lisix}{{}^6 \mathrm{Li}}
\newcommand{\Hefour}{{}^4 \mathrm{He}}
\newcommand{\Lisixnarrow}{{}^6\! \mathrm{Li}}
\newcommand{\Hefournarrow}{{}^4\! \mathrm{He}}
\newcommand{\taustau}{\tau_{\widetilde{\tau}_1}}
\newcommand{\TR}{T_{\mathrm{R}}}
\newcommand{\Tf}{T_{\mathrm{f}}}

\newcommand{\monetwo}{m_{1/2}}
\newcommand{\mzero}{m_{0}}
\newcommand{\tanb}{\tan{\beta}}
\newcommand{\mgut}{M_\mathrm{GUT}}
\newcommand{\Omegatp}{\Omega_{\widetilde{G}}^{\mathrm{TP}}}
\newcommand{\Omegantp}{\Omega_{\widetilde{G}}^{\mathrm{NTP}}}
\newcommand{\sigmarec}{\langle \sigma_{\mathrm{r}} v \rangle}
\newcommand{\sigmaC}{\langle \sigma_{\mathrm{C}} v \rangle}
\newcommand{\BS}{\mathrm{BS}}
\newcommand{\champ}{X^{\! -}}
\newcommand{\Ebind}{E_{\mathrm{b}}}

\begin{document}

\preprint{MPP-2007-115}
\preprint{arXiv:0710.2213}

\title{Implications of Catalyzed BBN in the CMSSM with Gravitino Dark Matter}

\author{Josef~Pradler}
\email{jpradler@mppmu.mpg.de}
\affiliation{Max-Planck-Institut f\"ur Physik, 
F\"ohringer Ring 6,
D--80805 Munich, Germany}
\author{Frank Daniel Steffen}
\email{steffen@mppmu.mpg.de}
\affiliation{Max-Planck-Institut f\"ur Physik, 
F\"ohringer Ring 6,
D--80805 Munich, Germany}

\begin{abstract} 
  We investigate gravitino dark matter scenarios in which the
  primordial $\Lisix$ production is catalyzed by bound-state formation
  of long-lived negatively charged particles $\champ$ with $\Hefour$.
  In the constrained minimal supersymmetric Standard Model (CMSSM)
  with the stau $\stau^-$ as the $\champ$,
  the observationally inferred bound on the primordial $\Lisix$
  abundance allows us to derive a rigid lower limit on the gaugino
  mass parameter for a standard cosmological history.
  This limit can have severe implications for supersymmetry searches
  at the Large Hadron Collider and for the reheating temperature after
  inflation.
\end{abstract}

%
%
\maketitle

\section{Introduction}
\label{sec:introduction}

Big Bang Nucleosynthesis (BBN) is a powerful tool to test physics
beyond the Standard Mo\-del.
Recently, it has been realized that the presence of heavy long-lived
negatively charged particles $\champ$ can have a substantial impact on
the primordial light element abundances via bound-state
formation%
~\cite{Pospelov:2006sc,Kohri:2006cn,Kaplinghat:2006qr,Cyburt:2006uv,Hamaguchi:2007mp,Bird:2007ge,Kawasaki:2007xb,Jittoh:2007fr,Jedamzik:2007cp}.
In particular, when $\champ$ and $\Hefour$ form Coulomb bound states,
$(\Hefour\champ)$, too much $\Lisix$ can be produced via the catalyzed
BBN (CBBN) reaction~\cite{Pospelov:2006sc}
\begin{align}
\label{eq:CBBN-reaction}
  (\Hefour\champ)+\mathrm{D} \rightarrow \Lisix + \champ .
\end{align}
The formation of $(\Hefour\champ)$ and hence the CBBN production of
$\Lisix$ becomes efficient at temperatures $T \sim 10\ \keV$, i.e., at
cosmic times $t> 10^3\ \seconds$ at which standard BBN (SBBN)
processes are already frozen out.  The observationally inferred bound
on the primordial $\Lisix$ abundance then restricts severely the
$\champ$ abundance at such times.

A long-lived $\champ$ may be realized if the gravitino is the lightest
supersymmetric particle (LSP).  In particular, it is reasonable to
consider gravitino LSP scenarios within the constrained minimal
supersymmetric Standard Model
(CMSSM)~\cite{Ellis:2003dn,Cerdeno:2005eu,Jedamzik:2005dh,Cyburt:2006uv,Pradler:2006hh}
in which the gaugino masses, the scalar masses, and the trilinear
scalar couplings are parameterized by their respective universal
values $\monetwo$, $\mzero$, and $A_0$ at the scale of grand
unification $\mgut \simeq 2\times 10^{16}\ \GeV$.  Within this
framework, the lighter stau $\stau$ is the lightest Standard Model
superpartner in a large region of the parameter space and thus a
well-motivated candidate for the next-to-lightest supersymmetric
particle (NLSP).  Since its couplings to the gravitino LSP are
suppressed by the (reduced) Planck scale, $\MP=2.4\times
10^{18}\,\GeV$, the stau will typically be long-lived for conserved
$\mathrm{R}$-parity%
\footnote{For the case of broken R-parity, see, e.g.~\cite{Buchmuller:2007ui}}
and thus $\stau^-$ can play the role of
$\champ$. 

In scenarios with conserved $\mathrm{R}$-parity, the gravitino LSP is
stable and a promisig dark matter candidate. Gravitinos can be
produced efficiently in thermal scattering of particles in the
primordial plasma. If the Universe, after inflation, enters the
radiation dominated epoch with a high reheating temperature $\TR$, the
resulting gravitino density $\Omegatp$ will contribute substantially
to the dark matter density
$\Omega_{\mathrm{dm}}$~\cite{Bolz:2000fu,Pradler:2006qh,Rychkov:2007uq}.

In this work we calculate the amount of $\Lisix$ produced
in~(\ref{eq:CBBN-reaction}) by following the treatment of
Ref.~\cite{Takayama:2007du}. In particluar, we employ a recent
state-of-the-art result for the CBBN reaction cross
reaction~\cite{Hamaguchi:2007mp}. The obtained
upper limit on the $\champ$ abundance from possible $\Lisix$
overproduction vanishes for sufficiently short $\tau_{\champ}$. This
allows us to extract a lower limit on the universal gaugino mass
parameter $\monetwo$ within minimal supergravity scenarios where the
gravitino is the LSP and the $\champ$ is the $\stau^-$
NLSP.\footnote{In this work we assume a standard cosmological history
  with a reheating temperature $\TR$ that exceeds the freeze-out
  temperature $\Tf$ of the $\stau$ NLSP.}
This limit leads directly to an upper bound on $\TR$ since $\Omegatp$
cannot exceed the observed dark matter density. The bounds on
$\monetwo$ and $\TR$ derived below depend on the gravitino mass but
are independent of the CMSSM parameters.

Before proceeding, let us comment on the present status of BBN
constraints on gravitino dark matter scenarios with a long-lived
charged slepton NLSP. In a recent ambitious
study~\cite{Jedamzik:2007cp} it is argued that bound-state formation
of $\champ$ with protons at $T \sim 1\ \keV$ might well reprocess
large fractions of the previously synthesized $\Lisix$.  This seems to
relax the bound on the $\champ$ abundance for $\tau_{\champ}>
10^6~\seconds$.  However, at present, the uncertainties in the
relevant nuclear reaction rates in~\cite{Jedamzik:2007cp} make it
difficult to decide whether a new cosmologically allowed region will
open up.  In this work we assume that this is not the case, in
particular, since the $^3$He/D constraint on electromagnetic energy
release~\cite{Sigl:1995kk} becomes severe in this region and excludes
stau lifetimes
$\tau_{\stau}\gtrsim10^6~\seconds$~\cite{Cerdeno:2005eu,Cyburt:2006uv,Kawasaki:2007xb,Jedamzik:2007cp}.
Then only the constraint from hadronic energy release on
D~\cite{Kawasaki:2004qu,Feng:2004mt,Cerdeno:2005eu,Jedamzik:2006xz,Steffen:2006hw}
can be slightly more severe than the one from catalyzed $^6$Li
production~\cite{Cyburt:2006uv,Steffen:2006wx,Pradler:2006hh,Kawasaki:2007xb}.
We neglect the D constraint in this work since it can only tighten the
bounds on $\monetwo$ and $\TR$ as can be seen, e.g., in Figs.~4~(b--d)
and~5 of Ref.~\cite{Pradler:2006hh}. For deriving conservative bounds
on $\monetwo$ and $\TR$, it is thus sufficient to consider the CBBN
reaction~(\ref{eq:CBBN-reaction}) exclusively.

\section{Catalyzed \boldmath{$\Lisix$} production}
\label{sec:catalyzed-bbn}

The following set of Boltzmann equations~\cite{Takayama:2007du}
describe the time evolution of bound-state (BS) formation of $\champ$
with $\Hefour$ and the associated evolution of the primordial light
elements involved:
\begin{subequations}
  \label{eq:boltzmann-eqs}
\begin{align}
 \label{eq:boltz-BS}
  \frac{d Y_{\BS} }{dt} 
  & = \sigmarec   s Y_\delta
  - \Gamma_{\champ}  Y_{\BS} 
  - \sigmaC s  Y_{\BS} Y_{\mathrm{D}} , 
  \\
  \frac{d Y_{\champ} }{dt} 
  & = 
  - \sigmarec   s Y_\delta
  - \Gamma_{\champ} Y_{\champ}  
  + \sigmaC s  Y_{\BS} Y_{\mathrm{D}} ,\\  
  \frac{d Y_{\Hefournarrow} }{dt} 
  & =  - \sigmarec  s Y_\delta
  +  \Gamma_{\champ} Y_{\BS} , \\ 
  \label{eq:boltz-Li}
  \frac{d Y_{\Lisixnarrow} }{dt} 
  & =  \sigmaC s  Y_{\BS} Y_{\mathrm{D}} ,
  \\
  \frac{d Y_{\mathrm{D}} }{dt} 
  & =  - \sigmaC s Y_{\BS} Y_{\mathrm{D}} .
\end{align}
\end{subequations}
Here we scale out the expansion of the Universe by defining the yield
$Y_i = n_i / s$ where $n_i$ is the number density of species $i$ and
$s = 2\pi^2\,g_{*S}\,T^3/45$ is the entropy density. In particular,
$Y_{\BS}$, $Y_{\champ}$, $Y_{\Hefournarrow}$, $Y_{\Lisixnarrow}$, and
$Y_{\mathrm{D}}$ denote the yields of the $(\Hefour\champ)$ bound
state, free $\champ$, free $\Hefour$, $\Lisix$ produced in CBBN, and
$\mathrm{D}$, respectively. The quantity $Y_\delta \equiv (Y_{\champ}
Y_{\Hefournarrow} - Y_{\BS} \widetilde{Y}_\gamma )$ parameterizes the
competition between recombination and photo-dissociation of bound
states. For the latter, one defines $\widetilde{Y}_\gamma =
\widetilde{n}_\gamma /s $ with~\cite{Kohri:2006cn}
\begin{align}
\label{eq:ngammaprime}
  \widetilde{n}_\gamma \equiv n_\gamma (E\!>\!\Ebind) 
  & = 
   \frac{n_{\gamma} \pi^2}{2 \zeta(3)} 
  \left( \frac{m_{\alpha}}{2 \pi T} \right)^{3/2}
  e^{-\Ebind /T},
\end{align}
where $\Ebind = 337.33\ \keV$~\cite{Hamaguchi:2007mp} is the
$(\Hefour\champ)$ binding energy and $n_\gamma = 2 \zeta(3) T^3 /
\pi^2$.  Furthermore, $\Gamma_{\champ} = \tau_{\champ}^{-1} $ denotes
the total decay width of $\champ$.

The CBBN reaction cross section for the process~(\ref{eq:CBBN-reaction})
has recently been computed with an advanced method from nuclear
physics~\cite{Hamaguchi:2007mp}
\begin{align}
  \label{eq:LiSigmav}
  \sigmaC = 2.37\times 10^8 \left( 1 - 0.34 T_9 \right) T_9^{-2/3}
  e^{-5.33 T_9^{-1/3}} 
\end{align}
which is given in units of $N_\mathrm{A}^{-1} \mathrm{cm}^3
\seconds^{-1} \mathrm{mole}^{-1}$ with $T_9$ denoting
the temperature in units of $10^9\ \mathrm{K}$.  The recombination
cross section of $\champ$ with $\Hefour$ is estimated
as~\cite{Kohri:2006cn}
\begin{align}
  \label{eq:sigmaREC}
  \sigmarec & = 
  \frac{2^9 \pi \alpha Z_{\alpha}^2 \sqrt{2 \pi}}{3
    e^4} \frac{\Ebind}{m_{\alpha}^2 \sqrt{m_{\alpha} T}}
\end{align}
with $m_\alpha = 3.73\ \GeV$~\cite{Yao:2006px} and $Z_{\alpha} = 2$.%
\footnote{Equation~(\ref{eq:sigmaREC}) assumes a radiative capture of
  $\champ$ into the 1S bound state of a point-like $\alpha$
  particle. We use, however, $E_\mathrm{b}$ obtained numerically
  in~\cite{Hamaguchi:2007mp} rather than the Bohr-like formula
  $E_{\mathrm{b}}^{0} \simeq Z_{\alpha}^2
   \alpha^2 m_{\alpha}/2 = 397\ \keV$;  $m_{\alpha}\ll
  m_{\champ}$.}
\begin{figure}
\includegraphics[width=0.45\textwidth,angle=0]{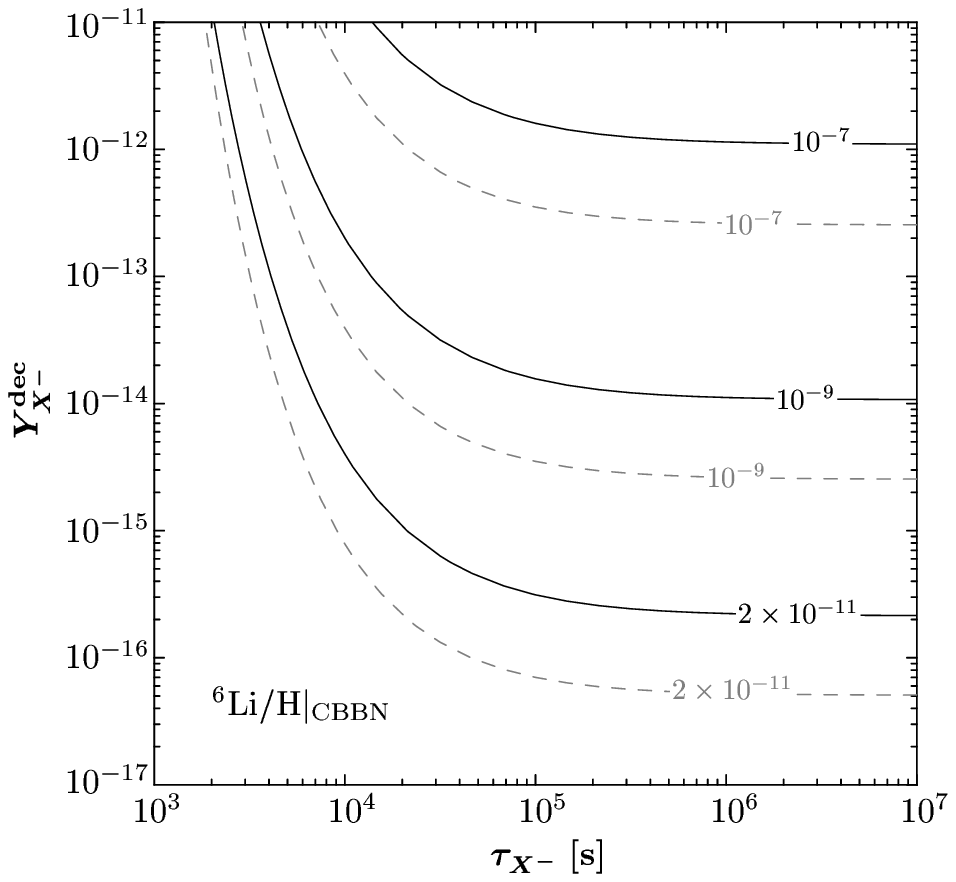}
\caption{Contour-lines of $\Lisix / \mathrm{H}$ produced in CBBN
  obtained by solving~(\ref{eq:boltzmann-eqs}) (solid lines) and by
  using the Saha type approximation for $Y_{\BS}$ instead of
  computing~(\ref{eq:boltz-BS}) (dashed lines).}
\label{fig:1}       
\end{figure}

We solve~(\ref{eq:boltzmann-eqs}) using as initial conditions the
respective $\champ$ yield prior to decay, $Y^{\mathrm{dec}}_{\champ}$,
and the SBBN output values of the computer
code~\texttt{PArthENoPE}~\cite{Pisanti:2007hk}: $Y_\mathrm{p} \equiv 4
n_{\Hefournarrow} / n_{\mathrm{b}} = 0.248 $, $\mathrm{D}/\mathrm{H}
=2.6\times 10^{-5}$, $\mathrm{\Lisix}/\mathrm{H} =1.14\times
10^{-14}$, and $n_{\mathrm{p}}/n_{\mathrm{b}} = 0.75$; furthermore,
$g_{*} = 3.36$ and $g_{*S} = 3.91$. While the variation of the
$\mathrm{D}$ and $\Hefour$ abundances from their SBBN values are
negligible, the catalyzed fusion of $\Lisix$ is substantial as shown
in Fig.~\ref{fig:1} by the contour-lines of $\Lisix/\mathrm{H} \equiv
Y_{\Lisix} s / n_{\mathrm{p}}$ (solid lines).
Contrasting with the observationally inferred upper limit on the
primordial $\Lisix$ abundance~\cite{Cyburt:2002uv},
\begin{align}
  \label{eq:Li-obs}
  \Lisix/\mathrm{H} |_{\mathrm{obs}} \lesssim 2\times 10^{-11},
\end{align}
one sees clearly that $\Lisix/\mathrm{H} |_{\mathrm{CBBN}}$ can be far
in excess. 

The dashed lines in Fig.~\ref{fig:1} show the solution of
(\ref{eq:boltzmann-eqs}) where instead of (\ref{eq:boltz-BS}) the Saha
type equation $Y_{\BS} = Y_{\Hefournarrow}Y_{\champ} /
\widetilde{Y}_\gamma $ is used as an approximation for the bound-state
abundance. The obtained overestimation of the $\Lisix$ abundance
demonstrates the importance of the use of the Boltzmann
equation~(\ref{eq:boltz-BS}). However, focusing on
$Y^{\mathrm{dec}}_{\champ}\gtrsim 10^{-14}$, we will read off the
relevant constraint in the region $\tau_{\champ} < 10^4\,\seconds$ in
which the slope of the $\Lisix$ contours is very steep. Therefore, the
use of (\ref{eq:boltz-BS}) instead of the Saha type equation is an
improvement on the conceptional side which has only a marginal effect
on the bounds to be derived. By the same token, and from the
comparison of our results with~\cite{Cyburt:2006uv}, we find that also
the destruction of $\Lisix$ due to $\champ$ decays affects those
bounds only marginally.

\section{Lower limit on \boldmath{$\monetwo$}}
\label{lowerlimitm12}
Applying the above results to gravitino dark matter scenarios with the
lighter stau $\stau$ as the NLSP, we now derive the conservative lower
limit on $\monetwo$.
The stau NLSP with a mass of $\mstau$ decouples from the primordial
plas\-ma with a typical yield of $Y^{\mathrm{dec}}_{\stau} \gtrsim
7\times 10^{-14} (\mstau / 100\ \GeV)$~\cite{Asaka:2000zh}.  With
$Y^{\mathrm{dec}}_{\champ} = Y^{\mathrm{dec}}_{\stau}/2$, we find from
Fig.~\ref{fig:1} that the amount of $\Lisix$ produced in CBBN can be
in agreement with (\ref{eq:Li-obs}) only for stau lifetimes of
\begin{align}
\label{eq:lifetimebound}
  \taustau = \tau_{\champ} \lesssim 5\times 10^3\;\seconds .
\end{align}

As can be seen from the supergravity prediction
\begin{align}
  \label{eq:taustau}
  \taustau 
  \simeq 
  \Gamma^{-1}(\stau \rightarrow \widetilde{G} \tau) 
  & =
  \frac {48 \pi \mgr^2 \MP^2}{ \mstau^{5} }
 \left(1-\frac{\mgr^2}{m_{\stau}^2}\right)^{-4} ,
\end{align}
the requirement (\ref{eq:lifetimebound}) implies a lower limit on the
splitting between $m_{\stau}$ and $\mgr$ provided $\mstau \lesssim
\mathcal{O}(1\ \TeV)$.  Because of this hierarchy, the factor
$(1-\mgr^2/\mstau^{2})^{-4}$ can be neglected in the following.

Let us now turn to the CMSSM. In the region in which~$\stau$ is the
NLSP, we find
\begin{align}
\label{eq:mstau-estimate}
  \mstau^2 \le 0.21 \monetwo^2 
\end{align}
by scanning over the following parameter range:
\begin{align*}
  \monetwo &= 0.1 - 6\ \TeV,\\ 
  \tanb &= 2-60, \\
 \mathrm{sgn\ \!}\mu &= \pm 1,\\
   -4 \mzero &< A_{0} < 4 \mzero
\end{align*}
with $\mzero$ as large as viable for a $\stau$ NLSP. Here $\tanb$
is the ratio of the two MSSM Higgs doublet vacuum expectation
values and $\mu$ the Higgsino mass parameter.%
\footnote{We employ \texttt{SPheno~2.2.3} \cite{Porod:2003um} to
  compute the low energy mass spectrum using $m_{\mathrm{t}} = 172.5\
  \GeV$ for the top quark mass. In addition, we use the Standard Model parameters
  $m_{\mathrm{b}}(m_{\mathrm{b}})^{\mathrm{\overline{MS}}} = 4.2\
  \GeV$,
  $\alpha_{\mathrm{s}}^{\mathrm{\overline{MS}}}(m_\mathrm{Z})=0.1172$,
  $\alpha_{\mathrm{em}}^{-1\mathrm{\overline{MS}}}(m_{\mathrm{Z}}) =
  127.932 $.}

For small left-right mixing, $\stau \simeq
\widetilde{\tau}_{\mathrm{R}}$, (\ref{eq:mstau-estimate}) can be
understood qualitatively from the estimate for the mass of the
right-handed stau $m_{\widetilde{\tau}_{\mathrm{R}}}$ near the
electroweak scale \cite{Martin:1993ft}
\begin{align}
 \label{eq:mstauR-estimate}
  m_{\widetilde{\tau}_{\mathrm{R}}}^2 \simeq  0.15
  \monetwo^2 + \mzero^2
  -\sin^2{\theta_{W}}m_{\mathrm{Z}}^2\cos{2\beta}\ .
\end{align}
since $\mzero^2 \ll \monetwo^2$ in a large part of the $\stau$ NLSP
region. In fact, (\ref{eq:mstau-estimate}) tends to be saturated for
larger $\mzero$, i.e., in the stau-neutralino-coannihilation region
where the mass of the lightest neutralino $m_{\neutralino}\simeq
\mstau$. This can be understood since the neutralino is bino-like in
this region so that $m_{\neutralino}^2 \simeq 0.18\monetwo^2$.%
\footnote{This estimate is relatively  independent of $\tanb$ and
  valid in the $\monetwo$ region in which also the LEP bound on the
  Higgs mass~\cite{Yao:2006px}, $m_{\mathrm{h}}>114.4\ \GeV$, is respected.  }
In the remaining part of the stau NLSP
region, smaller values of $\mstau$ 
satisfying, e.g., $\mstau^2 = 0.15\monetwo^2$ can easily be
found.

\begin{figure}
\includegraphics[width=0.45\textwidth,angle=0]{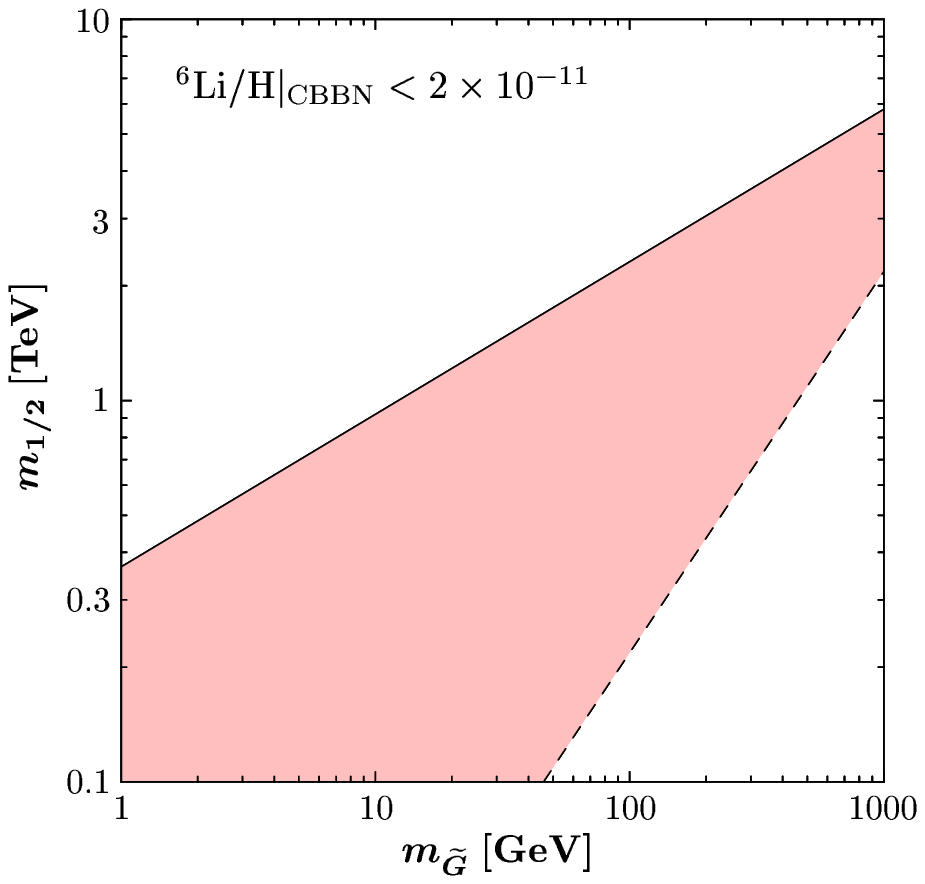}
\caption{The shaded region indicates cosmologically disfavored
  $\monetwo$ values. Below the dashed line, $\mgr\ge \mstau$ is
  possible.}
\label{fig:2}       
\end{figure}

To be on the conservative side, we set the stau NLSP mass $\mstau$ to
its maximum value at which (\ref{eq:mstau-estimate}) is saturated:
$\mstau^2 = 0.21\monetwo^2 $. Then,
constraint~(\ref{eq:lifetimebound}) together with~(\ref{eq:taustau})
yields
\begin{align}
  \label{eq:LowerLimitm12}
  \monetwo \ge 0.9\, \TeV \left( \frac{\mgr}{ 10\ \GeV}
  \right)^{2/5}  
\end{align}
which is shown in Fig.~\ref{fig:2}.
The shaded region is disfavored by~(\ref{eq:Li-obs}).
Below the dashed line, $\mgr\geq m_{\stau}$ is possible.

Since for a $\stau$ NLSP typically $ \mzero^2 \ll \monetwo^2 $, it is
the gaugino mass parameter $\monetwo$ which sets the scale for the low
energy superparticle spectrum. Thus, depending on $\mgr$, the
bound~(\ref{eq:LowerLimitm12}) implies rather high values of the
superparticle masses. This is particularly true for the masses of the
squarks and the gluino since their renormalization group running from
$\mgut$ to $Q\simeq \mathcal{O}(1\ \TeV)$ is dominated by
$M_3(Q)\simeq \monetwo \alpha_{\mathrm{s}}(Q) /
\alpha_{\mathrm{s}}(\mgut)$. Since these masses govern the size of the
total cross section for the production of superparticles at the Large
Hadron Collider (LHC), the cosmologically favored region for $\mgr
\gtrsim 10~\GeV$ is associated with a mass range that will be very
difficult to probe at the LHC.

Let us stress at this point that the bounds~(\ref{eq:lifetimebound})
and~(\ref{eq:LowerLimitm12}) and their severe implications for
phenomenology at the LHC are valid only for the assumed standard
cosmological history with $\TR>\Tf$ and the associated considered
values of $Y^{\mathrm{dec}}_{\stau}\gtrsim 7 \times 10^{-14}$.  For
example, non-standard entropy production after the thermal $\stau$
NLSP freeze out and before BBN might dilute the stau abundance prior
to decay.  Thereby, the $\monetwo$ limit can be
relaxed~\cite{Pradler:2006hh}. Also for the case of inflation with a
low reheating temperature, $\TR<\Tf$, one can obtain a stau abundance
prior to decay that respects the $\Lisix$ constraint even for
$\taustau \gtrsim 5\times 10^3\;\seconds$~\cite{Takayama:2007du}.
Thus, for a non-standard cosmolocial history, an observation of staus
with $\taustau \gtrsim 5\times 10^3\;\seconds$ and other CMSSM
phenomenology at the LHC could still be viable even in gravitino LSP
scenarios with $\mgr \gtrsim 10~\GeV$.

\section{Upper bound on \boldmath{$\TR$}}
\label{upperboundTR}
The amount of gravitinos produced in thermal scattering is sensitive
to the reheating temperature $\TR$ and to the masses of the gauginos
and hence to $\monetwo$~\cite{Pradler:2006qh}. For a standard
cosmological history, the associated gravitino density can be
approximated by%
\footnote{For a discussion on the definition of $\TR$, see Sec.~2 in
  Ref.~\cite{Pradler:2006hh}.}
\begin{align}
  \label{eq:omega-tp}
  \Omegatp h^2 \simeq 0.32 \Big( \frac{10\ \GeV}{\mgr} \Big) 
  \Big( \frac{\monetwo}{1\ \TeV} \Big)^2 \Big(
    \frac{\TR}{10^{8}\ \GeV} \Big) .
\end{align}
This follows from Eq.~(3) of Ref.~\cite{Pradler:2006qh}. Here we use
that the running gaugino masses $M_i$ associated with the gauge groups
$\mathrm{SU}(3)_{\mathrm{c}}$, $\mathrm{SU}(2)_{\mathrm{L}}$, and
$\mathrm{U}(1)_{\mathrm{Y}}$ satisfy $M_3:M_2:M_1 \simeq 3:1.6:1 $ at
a representative scale of $10^8\,\GeV$ at which we also evaluate the
respective gauge couplings. Furthermore, we only need to take into
account the production of the spin-$1/2$ components of the gravitino
since (\ref{eq:LowerLimitm12}) implies $M_i^2/3\mgr^2 \gg 1$ for $\mgr
\gtrsim1\ \GeV$.

For a given $\monetwo$, the reheating temperature $\TR$ is limited
from above because $\Omegatp$ cannot exceed the dark matter
density~\cite{Yao:2006px} $ \Omega_{\mathrm{dm}}^{3\sigma} h^2 = 0.105
^ {+0.021} _{-0.030}$ where $h$ is the Hubble constant in units of
$100~\mathrm{km}\ \mathrm{Mpc}^{-1}\seconds^{-1}$.  Requiring 
\begin{align}
  \Omegatp h^2 \leq 0.126
\end{align}
and using the derived lower bound (\ref{eq:LowerLimitm12}) allows us
to extract the conservative upper limit:
\begin{align}
  \label{eq:UpperLimitTR}
  \TR \lesssim 4.9\times 10^7 \ \GeV \left( \frac{\mgr}{10\ \GeV}
  \right) ^{1/5} .
\end{align}
This constraint is a slowly varying function of $\mgr$: $(\mgr/10\,
\GeV)^{1/5} = 0.6 - 2.5$ for $\mgr = 1\, \GeV - 1\, \TeV$.  Therefore,
(\ref{eq:UpperLimitTR}) poses a strong bound on $\TR$ for the natural
gravitino LSP mass range in gravity-mediated supersymmetry breaking
scenarios.

Note that the constraint~(\ref{eq:UpperLimitTR}) relies on thermal
gravitino production only. In addition, gravitinos are produced in
stau NLSP decays with the respective density
\begin{align}
  \Omegantp h^2 = \mgr Y^{\mathrm{dec}}_{\stau} s(T_0)
  h^2/\rho_{\mathrm{c}}\ ,
\end{align}
where $\rho_{\mathrm{c}} / [s(T_0) h^2] = 3.6\times 10^{-9}\
\GeV$~\cite{Yao:2006px}. While the precise value of $Y^{\mathrm{dec}}_{\stau}$ depends
on the concrete choice of the CMSSM parameters, the upper limit
(\ref{eq:UpperLimitTR}) can only become more stringent by taking
$\Omegantp$ into account.  For exemplary CMSSM scenarios, this can be
seen from the $(\monetwo, \mzero)$ planes shown in Figs.~4 and 5 of
Ref.~\cite{Pradler:2006hh}.%
\footnote{Gravitino production from inflaton decay can also be
  sub\-stan\-tial; see, e.g.,~\cite{Endo:2007sz,Asaka:2006bv}. This
  can further tighten the bound~(\ref{eq:UpperLimitTR}).}  These
figures illustrate that the severe limits (\ref{eq:LowerLimitm12}) and
(\ref{eq:UpperLimitTR}) are conservative bounds.

\section{Conclusion}
\label{sec:conclusion}
We have considered the catalysis of $\Lisix$ production in CMSSM
scenarios with the gravitino LSP and the stau NLSP. Within a standard
cosmological history, the calculated $\Lisix$ abundance drops below
the observational limit on primordial $\Lisix$ for $\tau_{\stau}
\lesssim 5\times 10^3\,\seconds$.
Taken at face value, we find that this constraint translates into a
lower limit $\monetwo \ge 0.9\, \TeV ( \mgr / 10\, \GeV )^{2/5}$ in
the entire natural region of the CMSSM parameter space.
This implies a conservative upper bound $\TR\lesssim 4.9\times
10^7\GeV( \mgr / 10\, \GeV )^{1/5}$. 
The bounds on $\monetwo$ and $\TR$ not only confirm our previous
findings~\cite{Pradler:2006hh} but are also independent of the
particular values of the CMSSM parameters for the considered $\stau$
NLSP abundances.
%
\begin{acknowledgements}
We are grateful to T.~Plehn, S.~Reinartz, and A.~Weber for valuable discussions.
\end{acknowledgements}

\noindent\emph{Note added:} After submission of this work, a
substantially revised version~(v3) of~\cite{Jedamzik:2007cp} together
with~\cite{Jedamzik:2007qk} appeared on the \texttt{arXiv}.  The
results of these works affect our limits only mildly.  Because of the
huge effect of~(\ref{eq:CBBN-reaction}) on the $\Lisix$ abundance, our
relatively simple treatment of CBBN is sufficient for our purposes.
This is also confirmed by a direct comparison of our data with Figs.~1
and~2 of the more elaborate CBBN treatment in~\cite{Jedamzik:2007qk}
for $B_{\mathrm{h}} \lesssim 3\times 10^{-3}$ ($\mstau\le 2.7\,\TeV$,
i.e., $\monetwo\le 6\,\TeV$) \cite{Steffen:2006hw} at the relevant
times of $t\simeq \mathrm{few}\times 10^{3}\,\seconds$.  For a given
$\Lisix/\mathrm{H}|_{\mathrm{obs}}$ bound, the effect on the
$\taustau$ limit is less than a factor of 1.5.  In addition, adopting
$\Lisix/\mathrm{H}|_{\mathrm{obs}} \lesssim 4\times 10^{-11}\
(2.7\times 10^{-10})$ as used in~\cite{Jedamzik:2007qk}, the numbers
in our Eqs.~(\ref{eq:lifetimebound}), (\ref{eq:LowerLimitm12}),
and~(\ref{eq:UpperLimitTR}) change respectively to $6\times 10^{3}\
(10^4)$, $0.87\ (0.78)$, and $5.3\times 10^{7}\ (6.5\times 10^{7})$.
Furthermore, by taking into account the uncertainties in the relevant
nuclear reaction rates, it is shown explicitly in Fig.~14 in v3
of~\cite{Jedamzik:2007cp} and in Fig.~5 in~\cite{Jedamzik:2007qk} that
cosmologically allowed regions for $\taustau\gtrsim 10^5\,\seconds$
are indeed extremely unlikely ($<1\%$) for $Y^{\mathrm{dec}}_{\stau}
\gtrsim 7\times 10^{-14} (\mstau / 100\ \GeV)$ even with
$f_{\mathrm{em}}$ as small as $3\times 10^{-2}$. Only with a finely
tuned $\mstau$--$\mgr$ degeneracy leading to $B_{\mathrm{h}}\to 0$ and
$f_{\mathrm{em}}\to 0$ can any bound on energy release and, in
particular, the one from $^3$He/D be evaded.


\begin{thebibliography}{33}
\expandafter\ifx\csname natexlab\endcsname\relax\def\natexlab#1{#1}\fi
\expandafter\ifx\csname bibnamefont\endcsname\relax
  \def\bibnamefont#1{#1}\fi
\expandafter\ifx\csname bibfnamefont\endcsname\relax
  \def\bibfnamefont#1{#1}\fi
\expandafter\ifx\csname citenamefont\endcsname\relax
  \def\citenamefont#1{#1}\fi
\expandafter\ifx\csname url\endcsname\relax
  \def\url#1{\texttt{#1}}\fi
\expandafter\ifx\csname urlprefix\endcsname\relax\def\urlprefix{URL }\fi
\providecommand{\bibinfo}[2]{#2}
\providecommand{\eprint}[2][]{\url{#2}}

\bibitem[{\citenamefont{Pospelov}(2007)}]{Pospelov:2006sc}
\bibinfo{author}{\bibfnamefont{M.}~\bibnamefont{Pospelov}},
  \bibinfo{journal}{Phys. Rev. Lett.} \textbf{\bibinfo{volume}{98}},
  \bibinfo{pages}{231301} (\bibinfo{year}{2007}), \eprint{hep-ph/0605215}.

\bibitem[{\citenamefont{Kohri and Takayama}(2007)}]{Kohri:2006cn}
\bibinfo{author}{\bibfnamefont{K.}~\bibnamefont{Kohri}} \bibnamefont{and}
  \bibinfo{author}{\bibfnamefont{F.}~\bibnamefont{Takayama}},
  \bibinfo{journal}{Phys. Rev.} \textbf{\bibinfo{volume}{D76}},
  \bibinfo{pages}{063507} (\bibinfo{year}{2007}), \eprint{hep-ph/0605243}.

\bibitem[{\citenamefont{Kaplinghat and Rajaraman}(2006)}]{Kaplinghat:2006qr}
\bibinfo{author}{\bibfnamefont{M.}~\bibnamefont{Kaplinghat}} \bibnamefont{and}
  \bibinfo{author}{\bibfnamefont{A.}~\bibnamefont{Rajaraman}},
  \bibinfo{journal}{Phys. Rev.} \textbf{\bibinfo{volume}{D74}},
  \bibinfo{pages}{103004} (\bibinfo{year}{2006}), \eprint{astro-ph/0606209}.

\bibitem[{\citenamefont{Cyburt et~al.}(2006)\citenamefont{Cyburt, Ellis,
  Fields, Olive, and Spanos}}]{Cyburt:2006uv}
\bibinfo{author}{\bibfnamefont{R.~H.} \bibnamefont{Cyburt}},
  \bibinfo{author}{\bibfnamefont{J.~R.} \bibnamefont{Ellis}},
  \bibinfo{author}{\bibfnamefont{B.~D.} \bibnamefont{Fields}},
  \bibinfo{author}{\bibfnamefont{K.~A.} \bibnamefont{Olive}}, \bibnamefont{and}
  \bibinfo{author}{\bibfnamefont{V.~C.} \bibnamefont{Spanos}},
  \bibinfo{journal}{JCAP} \textbf{\bibinfo{volume}{0611}}, \bibinfo{pages}{014}
  (\bibinfo{year}{2006}), \eprint{astro-ph/0608562}.

\bibitem[{\citenamefont{Hamaguchi et~al.}(2007)\citenamefont{Hamaguchi,
  Hatsuda, Kamimura, Kino, and Yanagida}}]{Hamaguchi:2007mp}
\bibinfo{author}{\bibfnamefont{K.}~\bibnamefont{Hamaguchi}},
  \bibinfo{author}{\bibfnamefont{T.}~\bibnamefont{Hatsuda}},
  \bibinfo{author}{\bibfnamefont{M.}~\bibnamefont{Kamimura}},
  \bibinfo{author}{\bibfnamefont{Y.}~\bibnamefont{Kino}}, \bibnamefont{and}
  \bibinfo{author}{\bibfnamefont{T.~T.} \bibnamefont{Yanagida}},
  \bibinfo{journal}{Phys. Lett.} \textbf{\bibinfo{volume}{B650}},
  \bibinfo{pages}{268} (\bibinfo{year}{2007}), \eprint{hep-ph/0702274}.

\bibitem[{\citenamefont{Bird et~al.}(2007)\citenamefont{Bird, Koopmans, and
  Pospelov}}]{Bird:2007ge}
\bibinfo{author}{\bibfnamefont{C.}~\bibnamefont{Bird}},
  \bibinfo{author}{\bibfnamefont{K.}~\bibnamefont{Koopmans}}, \bibnamefont{and}
  \bibinfo{author}{\bibfnamefont{M.}~\bibnamefont{Pospelov}}
  (\bibinfo{year}{2007}), \eprint{hep-ph/0703096}.

\bibitem[{\citenamefont{Kawasaki et~al.}(2007)\citenamefont{Kawasaki, Kohri,
  and Moroi}}]{Kawasaki:2007xb}
\bibinfo{author}{\bibfnamefont{M.}~\bibnamefont{Kawasaki}},
  \bibinfo{author}{\bibfnamefont{K.}~\bibnamefont{Kohri}}, \bibnamefont{and}
  \bibinfo{author}{\bibfnamefont{T.}~\bibnamefont{Moroi}},
  \bibinfo{journal}{Phys. Lett.} \textbf{\bibinfo{volume}{B649}},
  \bibinfo{pages}{436} (\bibinfo{year}{2007}), \eprint{hep-ph/0703122}.

\bibitem[{\citenamefont{Jittoh et~al.}(2007)}]{Jittoh:2007fr}
\bibinfo{author}{\bibfnamefont{T.}~\bibnamefont{Jittoh}} \bibnamefont{et~al.},
  \bibinfo{journal}{Phys. Rev.} \textbf{\bibinfo{volume}{D76}},
  \bibinfo{pages}{125023} (\bibinfo{year}{2007}), \eprint{0704.2914}.

\bibitem[{\citenamefont{Jedamzik}(2008{\natexlab{a}})}]{Jedamzik:2007cp}
\bibinfo{author}{\bibfnamefont{K.}~\bibnamefont{Jedamzik}},
  \bibinfo{journal}{Phys. Rev.} \textbf{\bibinfo{volume}{D77}},
  \bibinfo{pages}{063524} (\bibinfo{year}{2008}{\natexlab{a}}),
  \eprint{arxiv:0707.2070}.

\bibitem[{\citenamefont{Ellis et~al.}(2004)\citenamefont{Ellis, Olive, Santoso,
  and Spanos}}]{Ellis:2003dn}
\bibinfo{author}{\bibfnamefont{J.~R.} \bibnamefont{Ellis}},
  \bibinfo{author}{\bibfnamefont{K.~A.} \bibnamefont{Olive}},
  \bibinfo{author}{\bibfnamefont{Y.}~\bibnamefont{Santoso}}, \bibnamefont{and}
  \bibinfo{author}{\bibfnamefont{V.~C.} \bibnamefont{Spanos}},
  \bibinfo{journal}{Phys. Lett.} \textbf{\bibinfo{volume}{B588}},
  \bibinfo{pages}{7} (\bibinfo{year}{2004}), \eprint{hep-ph/0312262}.

\bibitem[{\citenamefont{Cerdeno et~al.}(2006)\citenamefont{Cerdeno, Choi,
  Jedamzik, Roszkowski, and Ruiz~de Austri}}]{Cerdeno:2005eu}
\bibinfo{author}{\bibfnamefont{D.~G.} \bibnamefont{Cerdeno}},
  \bibinfo{author}{\bibfnamefont{K.-Y.} \bibnamefont{Choi}},
  \bibinfo{author}{\bibfnamefont{K.}~\bibnamefont{Jedamzik}},
  \bibinfo{author}{\bibfnamefont{L.}~\bibnamefont{Roszkowski}},
  \bibnamefont{and} \bibinfo{author}{\bibfnamefont{R.}~\bibnamefont{Ruiz~de
  Austri}}, \bibinfo{journal}{JCAP} \textbf{\bibinfo{volume}{0606}},
  \bibinfo{pages}{005} (\bibinfo{year}{2006}), \eprint{hep-ph/0509275}.

\bibitem[{\citenamefont{Jedamzik et~al.}(2006)\citenamefont{Jedamzik, Choi,
  Roszkowski, and Ruiz~de Austri}}]{Jedamzik:2005dh}
\bibinfo{author}{\bibfnamefont{K.}~\bibnamefont{Jedamzik}},
  \bibinfo{author}{\bibfnamefont{K.-Y.} \bibnamefont{Choi}},
  \bibinfo{author}{\bibfnamefont{L.}~\bibnamefont{Roszkowski}},
  \bibnamefont{and} \bibinfo{author}{\bibfnamefont{R.}~\bibnamefont{Ruiz~de
  Austri}}, \bibinfo{journal}{JCAP} \textbf{\bibinfo{volume}{0607}},
  \bibinfo{pages}{007} (\bibinfo{year}{2006}), \eprint{hep-ph/0512044}.

\bibitem[{\citenamefont{Pradler and
  Steffen}(2007{\natexlab{a}})}]{Pradler:2006hh}
\bibinfo{author}{\bibfnamefont{J.}~\bibnamefont{Pradler}} \bibnamefont{and}
  \bibinfo{author}{\bibfnamefont{F.~D.} \bibnamefont{Steffen}},
  \bibinfo{journal}{Phys. Lett.} \textbf{\bibinfo{volume}{B648}},
  \bibinfo{pages}{224} (\bibinfo{year}{2007}{\natexlab{a}}),
  \eprint{hep-ph/0612291}.

\bibitem[{\citenamefont{{Buchm\"uller}
  et~al.}(2007)\citenamefont{{Buchm\"uller}, Covi, Hamaguchi, Ibarra, and
  Yanagida}}]{Buchmuller:2007ui}
\bibinfo{author}{\bibfnamefont{W.}~\bibnamefont{{Buchm\"uller}}},
  \bibinfo{author}{\bibfnamefont{L.}~\bibnamefont{Covi}},
  \bibinfo{author}{\bibfnamefont{K.}~\bibnamefont{Hamaguchi}},
  \bibinfo{author}{\bibfnamefont{A.}~\bibnamefont{Ibarra}}, \bibnamefont{and}
  \bibinfo{author}{\bibfnamefont{T.}~\bibnamefont{Yanagida}},
  \bibinfo{journal}{JHEP} \textbf{\bibinfo{volume}{03}}, \bibinfo{pages}{037}
  (\bibinfo{year}{2007}), \eprint{hep-ph/0702184}.

\bibitem[{\citenamefont{Bolz et~al.}(2001)\citenamefont{Bolz, Brandenburg, and
  {Buchm\"uller}}}]{Bolz:2000fu}
\bibinfo{author}{\bibfnamefont{M.}~\bibnamefont{Bolz}},
  \bibinfo{author}{\bibfnamefont{A.}~\bibnamefont{Brandenburg}},
  \bibnamefont{and}
  \bibinfo{author}{\bibfnamefont{W.}~\bibnamefont{{Buchm\"uller}}},
  \bibinfo{journal}{Nucl. Phys.} \textbf{\bibinfo{volume}{B606}},
  \bibinfo{pages}{518} (\bibinfo{year}{2001}), \eprint{hep-ph/0012052}.

\bibitem[{\citenamefont{Pradler and
  Steffen}(2007{\natexlab{b}})}]{Pradler:2006qh}
\bibinfo{author}{\bibfnamefont{J.}~\bibnamefont{Pradler}} \bibnamefont{and}
  \bibinfo{author}{\bibfnamefont{F.~D.} \bibnamefont{Steffen}},
  \bibinfo{journal}{Phys. Rev.} \textbf{\bibinfo{volume}{D75}},
  \bibinfo{pages}{023509} (\bibinfo{year}{2007}{\natexlab{b}}),
  \eprint{hep-ph/0608344}.

\bibitem[{\citenamefont{Rychkov and Strumia}(2007)}]{Rychkov:2007uq}
\bibinfo{author}{\bibfnamefont{V.~S.} \bibnamefont{Rychkov}} \bibnamefont{and}
  \bibinfo{author}{\bibfnamefont{A.}~\bibnamefont{Strumia}},
  \bibinfo{journal}{Phys. Rev.} \textbf{\bibinfo{volume}{D75}},
  \bibinfo{pages}{075011} (\bibinfo{year}{2007}), \eprint{hep-ph/0701104}.

\bibitem[{\citenamefont{Takayama}(2007)}]{Takayama:2007du}
\bibinfo{author}{\bibfnamefont{F.}~\bibnamefont{Takayama}}
  (\bibinfo{year}{2007}), \eprint{arXiv:0704.2785 [hep-ph]}.

\bibitem[{\citenamefont{Sigl et~al.}(1995)\citenamefont{Sigl, Jedamzik,
  Schramm, and Berezinsky}}]{Sigl:1995kk}
\bibinfo{author}{\bibfnamefont{G.}~\bibnamefont{Sigl}},
  \bibinfo{author}{\bibfnamefont{K.}~\bibnamefont{Jedamzik}},
  \bibinfo{author}{\bibfnamefont{D.~N.} \bibnamefont{Schramm}},
  \bibnamefont{and} \bibinfo{author}{\bibfnamefont{V.~S.}
  \bibnamefont{Berezinsky}}, \bibinfo{journal}{Phys. Rev.}
  \textbf{\bibinfo{volume}{D52}}, \bibinfo{pages}{6682} (\bibinfo{year}{1995}),
  \eprint{astro-ph/9503094}.

\bibitem[{\citenamefont{Kawasaki et~al.}(2005)\citenamefont{Kawasaki, Kohri,
  and Moroi}}]{Kawasaki:2004qu}
\bibinfo{author}{\bibfnamefont{M.}~\bibnamefont{Kawasaki}},
  \bibinfo{author}{\bibfnamefont{K.}~\bibnamefont{Kohri}}, \bibnamefont{and}
  \bibinfo{author}{\bibfnamefont{T.}~\bibnamefont{Moroi}},
  \bibinfo{journal}{Phys. Rev.} \textbf{\bibinfo{volume}{D71}},
  \bibinfo{pages}{083502} (\bibinfo{year}{2005}), \eprint{astro-ph/0408426}.

\bibitem[{\citenamefont{Feng et~al.}(2004)\citenamefont{Feng, Su, and
  Takayama}}]{Feng:2004mt}
\bibinfo{author}{\bibfnamefont{J.~L.} \bibnamefont{Feng}},
  \bibinfo{author}{\bibfnamefont{S.}~\bibnamefont{Su}}, \bibnamefont{and}
  \bibinfo{author}{\bibfnamefont{F.}~\bibnamefont{Takayama}},
  \bibinfo{journal}{Phys. Rev.} \textbf{\bibinfo{volume}{D70}},
  \bibinfo{pages}{075019} (\bibinfo{year}{2004}), \eprint{hep-ph/0404231}.

\bibitem[{\citenamefont{Jedamzik}(2006)}]{Jedamzik:2006xz}
\bibinfo{author}{\bibfnamefont{K.}~\bibnamefont{Jedamzik}},
  \bibinfo{journal}{Phys. Rev.} \textbf{\bibinfo{volume}{D74}},
  \bibinfo{pages}{103509} (\bibinfo{year}{2006}), \eprint{hep-ph/0604251}.

\bibitem[{\citenamefont{Steffen}(2006)}]{Steffen:2006hw}
\bibinfo{author}{\bibfnamefont{F.~D.} \bibnamefont{Steffen}},
  \bibinfo{journal}{JCAP} \textbf{\bibinfo{volume}{0609}}, \bibinfo{pages}{001}
  (\bibinfo{year}{2006}), \eprint{hep-ph/0605306}.

\bibitem[{\citenamefont{Steffen}(2007)}]{Steffen:2006wx}
\bibinfo{author}{\bibfnamefont{F.~D.} \bibnamefont{Steffen}},
  \bibinfo{journal}{AIP Conf. Proc.} \textbf{\bibinfo{volume}{903}},
  \bibinfo{pages}{595} (\bibinfo{year}{2007}), \eprint{hep-ph/0611027}.

\bibitem[{\citenamefont{Yao et~al.}(2006)}]{Yao:2006px}
\bibinfo{author}{\bibfnamefont{W.~M.} \bibnamefont{Yao}} \bibnamefont{et~al.}
  (\bibinfo{collaboration}{Particle Data Group}), \bibinfo{journal}{J. Phys.}
  \textbf{\bibinfo{volume}{G33}}, \bibinfo{pages}{1} (\bibinfo{year}{2006}).

\bibitem[{\citenamefont{Pisanti et~al.}(2007)}]{Pisanti:2007hk}
\bibinfo{author}{\bibfnamefont{O.}~\bibnamefont{Pisanti}} \bibnamefont{et~al.}
  (\bibinfo{year}{2007}), \eprint{arXiv:0705.0290 [astro-ph]}.

\bibitem[{\citenamefont{Cyburt et~al.}(2003)\citenamefont{Cyburt, Ellis,
  Fields, and Olive}}]{Cyburt:2002uv}
\bibinfo{author}{\bibfnamefont{R.~H.} \bibnamefont{Cyburt}},
  \bibinfo{author}{\bibfnamefont{J.~R.} \bibnamefont{Ellis}},
  \bibinfo{author}{\bibfnamefont{B.~D.} \bibnamefont{Fields}},
  \bibnamefont{and} \bibinfo{author}{\bibfnamefont{K.~A.} \bibnamefont{Olive}},
  \bibinfo{journal}{Phys. Rev.} \textbf{\bibinfo{volume}{D67}},
  \bibinfo{pages}{103521} (\bibinfo{year}{2003}), \eprint{astro-ph/0211258}.

\bibitem[{\citenamefont{Asaka et~al.}(2000)\citenamefont{Asaka, Hamaguchi, and
  Suzuki}}]{Asaka:2000zh}
\bibinfo{author}{\bibfnamefont{T.}~\bibnamefont{Asaka}},
  \bibinfo{author}{\bibfnamefont{K.}~\bibnamefont{Hamaguchi}},
  \bibnamefont{and} \bibinfo{author}{\bibfnamefont{K.}~\bibnamefont{Suzuki}},
  \bibinfo{journal}{Phys. Lett.} \textbf{\bibinfo{volume}{B490}},
  \bibinfo{pages}{136} (\bibinfo{year}{2000}), \eprint{hep-ph/0005136}.

\bibitem[{\citenamefont{Porod}(2003)}]{Porod:2003um}
\bibinfo{author}{\bibfnamefont{W.}~\bibnamefont{Porod}},
  \bibinfo{journal}{Comput. Phys. Commun.} \textbf{\bibinfo{volume}{153}},
  \bibinfo{pages}{275} (\bibinfo{year}{2003}), \eprint{hep-ph/0301101}.

\bibitem[{\citenamefont{Martin and Ramond}(1993)}]{Martin:1993ft}
\bibinfo{author}{\bibfnamefont{S.~P.} \bibnamefont{Martin}} \bibnamefont{and}
  \bibinfo{author}{\bibfnamefont{P.}~\bibnamefont{Ramond}},
  \bibinfo{journal}{Phys. Rev.} \textbf{\bibinfo{volume}{D48}},
  \bibinfo{pages}{5365} (\bibinfo{year}{1993}), \eprint{hep-ph/9306314}.

\bibitem[{\citenamefont{Endo et~al.}(2007)\citenamefont{Endo, Takahashi, and
  Yanagida}}]{Endo:2007sz}
\bibinfo{author}{\bibfnamefont{M.}~\bibnamefont{Endo}},
  \bibinfo{author}{\bibfnamefont{F.}~\bibnamefont{Takahashi}},
  \bibnamefont{and} \bibinfo{author}{\bibfnamefont{T.~T.}
  \bibnamefont{Yanagida}} (\bibinfo{year}{2007}), \eprint{arXiv:0706.0986
  [hep-ph]}.

\bibitem[{\citenamefont{Asaka et~al.}(2006)\citenamefont{Asaka, Nakamura, and
  Yamaguchi}}]{Asaka:2006bv}
\bibinfo{author}{\bibfnamefont{T.}~\bibnamefont{Asaka}},
  \bibinfo{author}{\bibfnamefont{S.}~\bibnamefont{Nakamura}}, \bibnamefont{and}
  \bibinfo{author}{\bibfnamefont{M.}~\bibnamefont{Yamaguchi}},
  \bibinfo{journal}{Phys. Rev.} \textbf{\bibinfo{volume}{D74}},
  \bibinfo{pages}{023520} (\bibinfo{year}{2006}), \eprint{hep-ph/0604132}.

\bibitem[{\citenamefont{Jedamzik}(2008{\natexlab{b}})}]{Jedamzik:2007qk}
\bibinfo{author}{\bibfnamefont{K.}~\bibnamefont{Jedamzik}},
  \bibinfo{journal}{JCAP} \textbf{\bibinfo{volume}{0803}}, \bibinfo{pages}{008}
  (\bibinfo{year}{2008}{\natexlab{b}}), \eprint{arxiv:0710.5153}.

\end{thebibliography}
\end{document}